\documentclass{llncs}
\usepackage{makeidx}

\usepackage[utf8]{inputenc}
\usepackage{amsmath,amssymb}
\usepackage{rotating}
\usepackage[all]{xy}
\usepackage{subfig}
\usepackage{multirow}
\usepackage{hyperref}
\usepackage{verbatim}
\usepackage{color}
\usepackage[usenames,dvipsnames]{xcolor}
\usepackage{footnote}

\newcommand{\mcb}[1]{\multicolumn{2}{c|}{#1}}

\makeatletter
\@ifundefined{lhs2tex.lhs2tex.sty.read}%
  {\@namedef{lhs2tex.lhs2tex.sty.read}{}%
   \newcommand\SkipToFmtEnd{}%
   \newcommand\EndFmtInput{}%
   \long\def\SkipToFmtEnd#1\EndFmtInput{}%
  }\SkipToFmtEnd

\newcommand\ReadOnlyOnce[1]{\@ifundefined{#1}{\@namedef{#1}{}}\SkipToFmtEnd}
\usepackage{amstext}
\usepackage{amssymb}
\usepackage{stmaryrd}
\DeclareFontFamily{OT1}{cmtex}{}
\DeclareFontShape{OT1}{cmtex}{m}{n}
  {<5><6><7><8>cmtex8
   <9>cmtex9
   <10><10.95><12><14.4><17.28><20.74><24.88>cmtex10}{}
\DeclareFontShape{OT1}{cmtex}{m}{it}
  {<-> ssub * cmtt/m/it}{}

\DeclareFontShape{OT1}{cmtt}{bx}{n}
  {<5><6><7><8>cmtt8
   <9>cmbtt9
   <10><10.95><12><14.4><17.28><20.74><24.88>cmbtt10}{}
\DeclareFontShape{OT1}{cmtex}{bx}{n}
  {<-> ssub * cmtt/bx/n}{}

\newcommand{\Conid}[1]{\mathit{#1}}
\newcommand{\Varid}[1]{\mathit{#1}}
\newcommand{\anonymous}{\kern0.06em \vbox{\hrule\@width.5em}}
\newcommand{\plus}{\mathbin{+\!\!\!+}}

\renewcommand{\leq}{\leqslant}

\usepackage{polytable}

\@ifundefined{mathindent}%
  {\newdimen\mathindent\mathindent\leftmargini}%
  {}%

\def\resethooks{%
  \global\let\SaveRestoreHook\empty
  \global\let\ColumnHook\empty}
\newcommand*{\savecolumns}[1][default]%
  {\g@addto@macro\SaveRestoreHook{\savecolumns[#1]}}
\newcommand*{\restorecolumns}[1][default]%
  {\g@addto@macro\SaveRestoreHook{\restorecolumns[#1]}}
\newcommand*{\aligncolumn}[2]%
  {\g@addto@macro\ColumnHook{\column{#1}{#2}}}

\resethooks

\newcommand{\onelinecommentchars}{\quad-{}- }
\newcommand{\commentbeginchars}{\enskip\{-}
\newcommand{\commentendchars}{-\}\enskip}

\newcommand{\visiblecomments}{%
  \let\onelinecomment=\onelinecommentchars
  \let\commentbegin=\commentbeginchars
  \let\commentend=\commentendchars}

\newcommand{\invisiblecomments}{%
  \let\onelinecomment=\empty
  \let\commentbegin=\empty
  \let\commentend=\empty}

\visiblecomments

\newlength{\blanklineskip}
\newcommand{\hsindent}[1]{\quad}
\let\hspre\empty
\let\hspost\empty

\EndFmtInput
\makeatother
\ReadOnlyOnce{polycode.fmt}%
\makeatletter

\newcommand{\hsnewpar}[1]%
  {{\parskip=0pt\parindent=0pt\par\vskip #1\noindent}}

\newcommand{\hscodestyle}{}

\newcommand{\sethscode}[1]%
  {\expandafter\let\expandafter\hscode\csname #1\endcsname
   \expandafter\let\expandafter\endhscode\csname end#1\endcsname}

  {\par\noindent
   \advance\leftskip\mathindent
   \hscodestyle
   \let\\=\@normalcr
   \let\hspre\(\let\hspost\)%
   \pboxed}%
  {\endpboxed\)%
   \par\noindent
   \ignorespacesafterend}

  {\hsnewpar\abovedisplayskip
   \advance\leftskip\mathindent
   \hscodestyle
   \let\hspre\(\let\hspost\)%
   \pboxed}%
  {\endpboxed%
   \hsnewpar\belowdisplayskip
   \ignorespacesafterend}

  {\hsnewpar\abovedisplayskip
   \advance\leftskip\mathindent
   \hscodestyle
   \let\\=\@normalcr
   \(\pboxed}%
  {\endpboxed\)%
   \hsnewpar\belowdisplayskip
   \ignorespacesafterend}

\newcommand{\plainhs}{\sethscode{plainhscode}}

\plainhs

  {\hsnewpar\abovedisplayskip
   \advance\leftskip\mathindent
   \hscodestyle
   \let\\=\@normalcr
   \(\parray}%
  {\endparray\)%
   \hsnewpar\belowdisplayskip
   \ignorespacesafterend}

  {\parray}{\endparray}

  {\(\parray}{\endparray\)}

\def\codeframewidth{\arrayrulewidth}
\RequirePackage{calc}

  {\parskip=\abovedisplayskip\par\noindent
   \hscodestyle
   \arrayrulewidth=\codeframewidth
   \tabular{@{}|p{\linewidth-2\arraycolsep-2\arrayrulewidth-2pt}|@{}}%
   \hline\framedhslinecorrect\\{-1.5ex}%
   \let\endoflinesave=\\
   \let\\=\@normalcr
   \(\pboxed}%
  {\endpboxed\)%
   \framedhslinecorrect\endoflinesave{.5ex}\hline
   \endtabular
   \parskip=\belowdisplayskip\par\noindent
   \ignorespacesafterend}

\newcommand{\framedhslinecorrect}[2]%
  {#1[#2]}

  {\(\def\column##1##2{}%
   \let\>\undefined\let\<\undefined\let\\\undefined
   \newcommand\>[1][]{}\newcommand\<[1][]{}\newcommand\\[1][]{}%
   \def\fromto##1##2##3{##3}%
   }{\) }%

  {\let\orighscode=\hscode
   \let\origendhscode=\endhscode
   \def\endhscode{\def\hscode{\endgroup\def\@currenvir{hscode}\\}\begingroup}
   \orighscode\def\hscode{\endgroup\def\@currenvir{hscode}}}%
  {\origendhscode
   \global\let\hscode=\orighscode
   \global\let\endhscode=\origendhscode}%

\makeatother
\EndFmtInput
\newcommand{\raiserot}[1]{{%
  \begin{sideways}%
    #1\;\;
  \end{sideways}%
  }%
}

\newcommand*{\arrows}{%
    \mathrel{\vcenter{\offinterlineskip
    \hbox{$\longleftarrow$}\vskip-1.8ex\hbox{$\longrightarrow$}}}}
    
\newcommand*{\rdasharrows}{%
    \mathrel{\vcenter{\offinterlineskip
    \hbox{$\longleftarrow$}\vskip-1.8ex\hbox{$\dashrightarrow$}}}}

\newcommand*{\ldasharrows}{%
    \mathrel{\vcenter{\offinterlineskip
    \hbox{$\dashleftarrow$}\vskip-1.8ex\hbox{$\longrightarrow$}}}}

\newcommand*{\dasharrows}{%
    \mathrel{\vcenter{\offinterlineskip
    \hbox{$\dashleftarrow$}\vskip-1.9ex\hbox{$\dashrightarrow$}}}}

\newcommand*{\larrow}{\longleftarrow}
\newcommand*{\rarrow}{\longrightarrow}
\newcommand*{\ldash}{\dashleftarrow}

\newcommand*{\dashes}{\dasharrows}
\newcommand*{\ldashes}{\ldasharrows}
\newcommand*{\rdashes}{\rdasharrows}

\begin{document}

\frontmatter          
\pagestyle{headings}  
\addtocmark{Properties of Bidirectional Transformations} 

\mainmatter              
%
\title{A Generic Scheme and Properties of Bidirectional Transformations}
\titlerunning{A Generic Scheme and Properties of Bidirectional Transformations}  
%
\author{Hugo Pacheco\inst{1} \and Nuno Macedo\inst{1} \and Alcino Cunha\inst{1} \and Janis Voigtl\"{a}nder\inst{2}}
\authorrunning{Pacheco \and Macedo \and Cunha \and Voigtl\"{a}nder}   
%
\tocauthor{Hugo Pacheco, Nuno Macedo, Alcino Cunha, Janis Voigtl\"{a}nder}
\institute{HASLab / INESC TEC \& Universidade do Minho, Braga, Portugal\\
\and
Institute for Computer Science, University of Bonn, Germany\\
}

\maketitle              

\begin{abstract}
  The recent rise of interest in bidirectional transformations (BXs)
  has led to the development of many BX frameworks, originating in
  diverse computer science disciplines.  From a user perspective,
  these frameworks vary significantly in both interface and
  predictability of the underlying bidirectionalization technique.  In
  this paper we start by presenting a generic BX scheme that can be
  instantiated to different concrete interfaces, by plugging-in the
  desired notion of update and traceability. Based on that scheme, we
  then present several desirable generic properties that may
  characterize a BX framework, and show how they can be instantiated
  to concrete interfaces. This generic presentation is useful when
  exploring the BX design space: it might help developers when designing new
  frameworks and end-users when comparing existing ones. We support
  the latter claim, by applying it in a comparative survey of popular
  existing BX frameworks.
\end{abstract}

\section{Introduction}
\label{sec:intro}

Bidirectional transformations (BXs) are a ``mechanism for maintaining the consistency of two (or more) related sources of information''~\cite{GRACE:09}. The challenge of writing BXs has been long known in the database community since the seminal studies on view-update translation by Bancilhon and Spyratos~\cite{Bancilhon:81} and Dayal and Bernstein~\cite{Dayal:82}.
More recently, the pioneering work of Foster et al. on combinatorial languages for BXs~\cite{Foster:07} has recast a lot of attention towards this challenge, and given the impulse to the birth of a new research field on BXs, uniting researchers from diverse computer science communities including programming languages, model-driven engineering and databases.
In the last ten years, this burgeoning interest in BXs has led to the proposal of a vast number of approaches~\cite{GRACE:09,Schurr:2008}, inspired by different visions of the problem and motivated by different contexts where the need for bidirectionality arises.

In face of the multitude and diversity of existing approaches, each tool has been tailored to answer the challenges of its particular bidirectional scenarios, and has evolved to support different formal properties and specification styles that best suit its needs. Therefore, many of the fundamental problems of the field are not yet well established, mostly due to the non-existence of a universal classifying system and to the lack of common theoretical grounds between many of the approaches. 
This makes it hard to compare the various solutions, to understand precisely their advantages and limitations and provide effective criteria for assessing progress in the field.

To move forward, this pressing unification need has been gaining voice in the BX subcommunities, and such a maturation effort has been slowly undergoing through a series of seminars and workshops, displayed in publications such as~\cite{GRACE:09,HSST11,BX12,Terwilliger:2012}.
Unfortunately, despite such significant effort, the community is still far from reaching a consensus on the terminology and properties that are desirable and satisfiable by BX tools.
For example, some BX programming languages satisfy specific properties that are hard to correlate with the properties satisfied by other languages or whose practical implications on the expressiveness and behavior of the corresponding BXs are hard to understand. Moreover, although some properties are already well understood in the databases or programming languages communities, their meaning and applicability in the MDE community remains unclear. On the other side, the specification style and deployment level of most BX programming languages are still not adequate for tackling real-world scenarios currently supported by model-driven BX approaches.

In this paper, we propose a generic BX scheme in which concrete interfaces can be instantiated by choosing the desired representation of the update and traceability information. On top of that scheme, we propose a set of generic semantic properties that embody the desirable bidirectional behavior of the transformations.
In addition, we walk through a number of
BX frameworks and instantiate them in our scheme by incrementally exploring its design space, and the expected properties naturally emerge from such exercise.
As a first attempt to validate our generic BX scheme, we present a comparative survey of up to 40 existing BX approaches emerging from diverse BX subcommunities, classified according to their interface and semantic properties. Besides providing an insightful high-level picture of the state-of-art of the field of BXs, it raises interesting questions about the key design features for classifying BXs.

Section~\ref{sec:framework} presents our generic scheme and explores the instantiations of some popular frameworks (namely mappings, lenses, maintainers, trigonal systems, edit lenses and symmetric delta-lenses). Section~\ref{sec:properties} presents the generic bidirectional properties which are then instantiated for the frameworks enumerated above, while Section~\ref{sec:survey-existing-bx} presents a first effort to survey existing BX techniques under the proposed axes.
Sections~\ref{section:relwork} and~\ref{section:conclusion} discuss related work on other classification efforts 
and set forth the path towards a more complete survey on the design space of BXs.

\section{Scheme}
\label{sec:framework}

We begin by clarifying what we mean by BX. The goal of a BX between \ensuremath{\Conid{A}} and \ensuremath{\Conid{B}} is to enforce 
consistency between values of types \ensuremath{\Conid{A}} and \ensuremath{\Conid{B}}. The term ``type''
should be understood as a broad placeholder for type, schema or
metamodel, with ``value'' denoting value, instance, or model,
accordingly. Consistency recovery is achieved by means of two
transformations \ensuremath{\mathsf{to}} and \ensuremath{\mathsf{from}} whose purpose is, respectively,
to propagate \ensuremath{\Conid{A}} updates into consistent \ensuremath{\Conid{B}} updates and vice-versa.
Some BX frameworks derive the two transformations from an explicitly declared \emph{consistency
  relation} \ensuremath{\mathsf{R}\subseteq\Conid{A}\;\!\!\times\!\!\;\Conid{B}} between both types. Often, conversely, the consistency relation is actually
expressed in terms of the underlying transformations. In these cases, usually one of the transformations
must be specified by the user, the opposite one being derived from
it. In some rare cases the consistency relation is an implicit notion of the system. 
Table~\ref{tab:consistency} summarizes these options.

\begin{table}[!t]
\begin{center}
\begin{tabular}{|l|c|p{8cm}|}
  \hline
  \textbf{Explicit} & E & There is an explicitly declared consistency relation. \\
  \hline
  \textbf{Transformation} & T & The consistency relation is one of the transformations.\\
  \hline
  \textbf{Implicit} & I & The consistency relation is implicit.\\
  \hline
\end{tabular}
\end{center}
\caption{Consistency Relation.}
\label{tab:consistency}
\vspace{-.2in}
\end{table}

This definition precludes some frameworks sometimes said to also be
BX. That is the case of frameworks with general synchronization
procedures that recover consistency between values that were updated
concurrently.
Notice that our goal is not to give a definitive definition of
what is BX (thus rejecting such frameworks as not being BX), but just to
clarify and limit the scope of this~paper.

We depict the two application scenarios of a BX between \ensuremath{\Conid{A}} and \ensuremath{\Conid{B}} as follows:
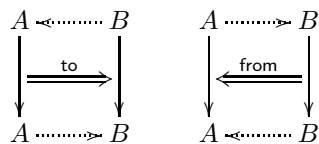
\begin{displaymath}
  \xymatrixrowsep{1pc}
  \xymatrix{
    A \ar[dd] & B \ar@{..>}[l] \ar[dd]\\
    \ar@{=>}[r]^{\ensuremath{\mathsf{to}}} &  \\
    A \ar@{..>}[r] & B 
  }
  \qquad
  \xymatrix{
    A \ar@{..>}[r] \ar[dd] & B \ar[dd]\\
    & \ar@{=>}[l]_{\ensuremath{\mathsf{from}}}  \\
    A & \ar@{..>}[l] B 
  }
\end{displaymath}
In the left scenario, we first have \ensuremath{\Varid{a}\mathbin{:}\Conid{A}} and \ensuremath{\Varid{b}\mathbin{:}\Conid{B}} that are
somehow consistent. The objective of \ensuremath{\mathsf{to}} is to transform an update on
\ensuremath{\Varid{a}} into an update on \ensuremath{\Varid{b}} such that consistency is restored. Updates
are depicted using solid arrows (although they are not necessarily functions, as clarified in Section~\ref{sec:update}).
The transformation \ensuremath{\mathsf{to}} may also
receive some extra information concerning the system state prior to
the update on \ensuremath{\Varid{a}}. Namely, it might have access to some trace
information testifying how \ensuremath{\Varid{a}} and \ensuremath{\Varid{b}} were consistent. Such
\emph{traceability} is depicted using a dotted arrow, and when not
trivially derived from the updated values must also be returned by
\ensuremath{\mathsf{to}}. The right scenario is dual.

In general, transformations \ensuremath{\mathsf{to}} and \ensuremath{\mathsf{from}} can be typed as follows:
\begin{hscode}\SaveRestoreHook
\column{B}{@{}>{\hspre}l<{\hspost}@{}}%
\column{7}{@{}>{\hspre}l<{\hspost}@{}}%
\column{E}{@{}>{\hspre}l<{\hspost}@{}}%
\>[B]{}\mathsf{to}{}\<[7]%
\>[7]{}\mathbin{:}\overrightarrow{\mathsf{U}}(\Conid{A})\;\!\!\times\!\!\;\overleftarrow{\mathsf{T}}(\Conid{A},\Conid{B})\to \overleftarrow{\mathsf{U}}(\Conid{B})\;\!\!\times\!\!\;\overrightarrow{\mathsf{T}}(\Conid{A},\Conid{B}){}\<[E]%
\\
\>[B]{}\mathsf{from}{}\<[7]%
\>[7]{}\mathbin{:}\overleftarrow{\mathsf{U}}(\Conid{B})\;\!\!\times\!\!\;\overrightarrow{\mathsf{T}}(\Conid{A},\Conid{B})\to \overrightarrow{\mathsf{U}}(\Conid{A})\;\!\!\times\!\!\;\overleftarrow{\mathsf{T}}(\Conid{A},\Conid{B}){}\<[E]%
\ColumnHook
\end{hscode}\resethooks
Here, \ensuremath{\overrightarrow{\mathsf{U}}} is a parameterized type constructor that denotes the type
of \ensuremath{\Conid{A}} updates that \ensuremath{\mathsf{to}} propagates, and \ensuremath{\overrightarrow{\mathsf{T}}} a type constructor that
denotes the type of the traceability \ensuremath{\mathsf{to}} should produce (to be
received by \ensuremath{\mathsf{from}}). Dually, we have \ensuremath{\overleftarrow{\mathsf{U}}} and \ensuremath{\overleftarrow{\mathsf{T}}} for \ensuremath{\mathsf{from}}. As we
will see in the next sections, the interface of existing (and
potential) BX frameworks can be obtained by giving concrete
definitions for these type constructors.
We should also clarify at this point that by transformation we mean a
\emph{partial function}. Frameworks differ on the degree of
totality of both transformations, which will be characterized by a specific
property (Section
\ref{sec:properties}).

This formal characterization encompasses both \emph{symmetric} and
\emph{asymmetric} frameworks (see Table~\ref{tab:symmetry}).
Asymmetric frameworks are often
biased towards transformation scenarios where one of the types is
``larger'' and contains more information than the other, whereas
symmetric frameworks are more balanced and tend to consider that both
types contain roughly the same information or that each may contain
information not present in the other.

\begin{table}[!t]
\begin{center}
\begin{tabular}{|l|c|p{9cm}|}
  \hline
  \textbf{Symmetric} & S & The update propagation nature is the same in both directions. We have equal definitions for \ensuremath{\overrightarrow{\mathsf{U}}} and \ensuremath{\overleftarrow{\mathsf{U}}}, for \ensuremath{\overrightarrow{\mathsf{T}}} and \ensuremath{\overleftarrow{\mathsf{T}}}, and similar laws for both \ensuremath{\mathsf{to}} and \ensuremath{\mathsf{from}}. \\
  \hline
  \textbf{Asymmetric} & A & The update propagation nature is different in both directions. This may lead to different definitions for \ensuremath{\overrightarrow{\mathsf{U}}} and \ensuremath{\overleftarrow{\mathsf{U}}}, for \ensuremath{\overrightarrow{\mathsf{T}}} and \ensuremath{\overleftarrow{\mathsf{T}}}, and different laws for \ensuremath{\mathsf{to}} and \ensuremath{\mathsf{from}}. \\
  \hline
\end{tabular}
\end{center}
\caption{Symmetry.}
\label{tab:symmetry}
\vspace{-.2in}
\end{table}

For each instance of the arrows in the above diagrams it may be useful to
reason about its source and target values.
We will use (overloaded)
operators \ensuremath{ \delta } and \ensuremath{ \rho } to denote them. 
Namely, for each update constructor \ensuremath{\overrightarrow{\mathsf{U}}}
(and dually for \ensuremath{\overleftarrow{\mathsf{U}}}), we have
\ensuremath{ \delta \mathbin{:}\overrightarrow{\mathsf{U}}(\Conid{A})\to \Conid{A}} and \ensuremath{ \rho \mathbin{:}\overrightarrow{\mathsf{U}}(\Conid{A})\to \Conid{A}} that, given an update, denote the value in
its pre- and post-state, respectively. 
Similarly for traceabilities,
\ensuremath{ \delta \mathbin{:}\overrightarrow{\mathsf{T}}(\Conid{A},\Conid{B})\to \Conid{A}} and \ensuremath{ \rho \mathbin{:}\overrightarrow{\mathsf{T}}(\Conid{A},\Conid{B})\to \Conid{B}} denote the source and
target values related by an instance of \ensuremath{\overrightarrow{\mathsf{T}}(\Conid{A},\Conid{B})}, respectively. 
\ensuremath{\overleftarrow{\mathsf{T}}} traceabilities
are seen in the other direction, so \ensuremath{ \delta \mathbin{:}\overleftarrow{\mathsf{T}}(\Conid{A},\Conid{B})\to \Conid{B}} and
\ensuremath{ \rho \mathbin{:}\overleftarrow{\mathsf{T}}(\Conid{A},\Conid{B})\to \Conid{A}}. 
We will denote updates and traceabilities by bold
characters (\ensuremath{{\textbf {\textit {\textrm a}}},{\textbf {\textit {\textrm b}}},{\textbf {\textit {\textrm r}}},{\textbf {\textit {\textrm s}}},\mathbin{...}}) in contrast to non-bold characters
for the values (\ensuremath{\Varid{a},\Varid{b},\mathbin{...}}) returned by these operators.
Although it is assumed that every arrow has a pre- and post-state value, that does not mean that they are retrieved directly from the update representation (that information may not even be present in the constructor).
When a transformation \ensuremath{\mathsf{to}\;({\textbf {\textit {\textrm a}}},{\textbf {\textit {\textrm s}}})\mathrel{=}({\textbf {\textit {\textrm b}}},{\textbf {\textit {\textrm r}}})} occurs, updates and
traceabilities must agree on the respective sources and targets. 
This can be captured by the following properties, coined
\emph{incidence conditions} in~\cite{Diskin:2011}.
\begin{align}
    \label{eq:incidence}
    \ensuremath{ \delta {\textbf {\textit {\textrm a}}}\mathrel{=}\rho {\textbf {\textit {\textrm s}}}\quad\quad \delta {\textbf {\textit {\textrm b}}}\mathrel{=} \delta {\textbf {\textit {\textrm s}}}\quad\quad\rho {\textbf {\textit {\textrm a}}}\mathrel{=} \delta {\textbf {\textit {\textrm r}}}\quad\quad\rho {\textbf {\textit {\textrm b}}}\mathrel{=}\rho {\textbf {\textit {\textrm r}}}}
\end{align}
The laws for \ensuremath{\mathsf{from}} are dual. These properties enable the retrieval of states even if not explicitly present in the constructor. For instance, an update \ensuremath{{\textbf {\textit {\textrm a}}}} may not have information about its pre-state, but it could for instance be retrieved from the traceability \ensuremath{{\textbf {\textit {\textrm s}}}}, since \ensuremath{ \delta {\textbf {\textit {\textrm a}}}\mathrel{=}\rho {\textbf {\textit {\textrm s}}}}.
We also assume that the input traceability truly testifies the consistency relation,
i.e., \ensuremath{(\rho {\textbf {\textit {\textrm s}}}, \delta {\textbf {\textit {\textrm s}}})\;\!\in\! \;\mathsf{R}}, which will be denoted by \ensuremath{{\textbf {\textit {\textrm s}}}\;\!\in\! \;\mathsf{R}}. Due
to the incidence conditions, \ensuremath{\rho {\textbf {\textit {\textrm s}}}} and \ensuremath{ \delta {\textbf {\textit {\textrm s}}}} can instead be accessed by \ensuremath{ \delta {\textbf {\textit {\textrm a}}}} and \ensuremath{ \delta {\textbf {\textit {\textrm b}}}},
respectively, whenever they are not present in traceability.

\subsection{Update representation}
\label{sec:update}

One of the main axes distinguishing existing frameworks is
\emph{update representation}, i.e., what are the concrete definitions
of \ensuremath{\overrightarrow{\mathsf{U}}} and \ensuremath{\overleftarrow{\mathsf{U}}} in the above generic scheme. Some possible
definitions are presented in Table~\ref{tab:update}.

\begin{table}[!t]
\begin{center}
\begin{tabular}{|l|c|p{7.5cm}|}
  \hline
  \parbox[t]{4cm}{\textbf{Post-state} \\\ensuremath{\overrightarrow{\mathsf{U}}(\Conid{A})\mathrel{=}\Conid{A}}} & S & An update is represented only by the post-state.\\ 
  \hline
  \parbox[t]{4cm}{\textbf{Both states} \\ \ensuremath{\overrightarrow{\mathsf{U}}(\Conid{A})\mathrel{=}\Conid{A}\;\!\!\times\!\!\;\Conid{A}}} & \ensuremath{\mathbb{S}} &The update is represented by the pre- and the post-state. \\
  \hline
  \parbox[t]{4cm}{\textbf{Delta} \\ \ensuremath{\overrightarrow{\mathsf{U}}(\Conid{A})\mathrel{=}\Conid{A}\;\!\!\times\!\!\;\Conid{A}\;\!\!\times\!\!\;\mathsf{D}(\Conid{A},\Conid{A})}} & D & Beside the pre- and the post-state, an update representation also comes with a \emph{sameness relation} stating which components of both are conceptually the same. \\
  \hline
  \parbox[t]{4cm}{\textbf{Edit} \\ \ensuremath{\overrightarrow{\mathsf{U}}(\Conid{A})\mathrel{=}\mathsf{O}(\Conid{A})^\star}} & E & An update is represented by the sequence of edit operations that was performed. \\
  \hline
  \parbox[t]{4cm}{\textbf{State + Edit} \\ \ensuremath{\overrightarrow{\mathsf{U}}(\Conid{A})\mathrel{=}\Conid{A}\;\!\!\times\!\!\;\mathsf{O}(\Conid{A})^\star}} & \ensuremath{\mathbb{E}} & An update is represented by the pre-state and the sequence of edit operations that was performed. \\
  \hline
  \parbox[t]{4cm}{\textbf{Function} \\ \ensuremath{\overrightarrow{\mathsf{U}}(\Conid{A})\mathrel{=}\Conid{A}\to \Conid{A}}} & F & An update is represented by a semantic value that models it as an endo-function. \\
  \hline
\end{tabular}
\end{center}
\caption{Update.}
\label{tab:update}
\vspace{-.2in}
\end{table}

Frameworks that fall within the first two categories are usually known
as \emph{state-based}, since only the value in the post- (and
sometimes also the pre-) state of an update is considered. When some
knowledge of the exact changes that were (or have to be) performed in
an update is represented, we have an \emph{operation-based}
framework. A possible way to represent such changes is via a
\emph{sameness relation} tracing back components of the post-state to
corresponding components in the pre-state. In Table~\ref{tab:update}
we encapsulate the concrete definition of such a sameness relation in
the parameterized type constructor \ensuremath{\mathsf{D}}, since it depends on the
domains involved in the transformation\footnote{\ensuremath{\mathsf{D}(\Conid{A},\Conid{B})} should not
  be confused with the set of all binary relations between \ensuremath{\Conid{A}} and
  \ensuremath{\Conid{B}}, which is denoted by \ensuremath{\mathcal{P}(\Conid{A}\;\!\!\times\!\!\;\Conid{B})}.}. Notice that a sameness
relation carries more information than the conjunction of pre- and
post-state alone: for example, if a component is deleted and a new one
inserted with the same content, these components would be
unconnected in the sameness relation, but indistinguishable
otherwise. A sequence of edit operations is another possible
representation for updates. Again, since the set of edit operations
supported by each type varies, we abstract away its concrete
definition in the parameterized type constructor \ensuremath{\mathsf{O}}. Another alternative
is to store the pre-state along with the edit-sequence, in which case
the post-state could also be calculated by applying the edit-sequence to
the pre-state. When updates
are performed programmatically, one might only have access to the
executable that performed them instead of a syntactic
representation. This information might still be exploited by
frameworks implemented on top of semantic BX techniques.

Although the extra knowledge in operation-based frameworks can lead to
BXs that satisfy more precise properties (see discussions
in~\cite{Diskin:2011,Hu:04journal}), they usually
demand a tight coupling with applications, so that they can track the
changes that characterize an update. Transforming updates onto updates
also makes BXs more natural with
incrementality~\cite{Giese:2006} (rather than recomputing a
new model when the correlated model changes, only a small ``delta'' is
propagated).  State-based frameworks, on the other hand, are more
flexible and support more usage scenarios, like integration with
off-the-shelf applications that have not been designed with
bidirectionality in mind, and are moreover less sensitive to ``noise''
in the updates.  The distinction between state- and operation-based
approaches is not always obvious. Hybrid approaches may build a
state-based system with a richer operation-based core.
As they discard all update information, some
model differencing procedure is required to infer new hypothetical
update operations. Similarly, an incremental system can have a simple
state-based core, but keep track of operations merely as an
optimization, to exploit the locality of updates.

\subsection{Traceability representation}

Likewise to update representation, Table~\ref{tab:traceability}
presents possible definitions for \emph{traceability representation},
i.e., the concrete definitions of \ensuremath{\overrightarrow{\mathsf{T}}} and \ensuremath{\overleftarrow{\mathsf{T}}} in the above generic
scheme. In the first category we have frameworks without any trace
information. This is a very limiting scenario, since only the to be translated
update itself is known when attempting to recover consistency. Not
even information about the current state of the opposite domain is
known: in a state-based framework this means that only ``fresh''
values must be produced, ruling out any sort of incremental
updating. In the second category traceability amounts precisely to the
value of the opposite domain. This is the case, for example, of
frameworks that tackle the view-update problem, and that require (specifically, \ensuremath{\mathsf{from}} requires) access
to the (previous) value of the source (\ensuremath{\Conid{A}}) to fetch information not
recoverable from the view (\ensuremath{\Conid{B}}).
Traceability can also be represented by
means of a \emph{complement}: \ensuremath{\mathsf{C}(\Conid{A},\Conid{B})} is a parameterized type
constructor that encapsulates the complement of a type \ensuremath{\Conid{A}} with
respect to another type \ensuremath{\Conid{B}}. Essentially, elements of \ensuremath{\mathsf{C}} will contain
some of the components of one (or both) value(s) not present in the other. Finally, we can
also use a sameness relation to trace the execution of a previous
update translation, pinpointing the exact pairs of components in the
source and target that testify the consistency between them.

Sometimes the returned traceability is redundant and can be computed
from the remaining available information. A possible instantiation where this occurs is when \ensuremath{\overrightarrow{\mathsf{U}}\mathrel{=}\text{S}} and \ensuremath{\overrightarrow{\mathsf{T}}\mathrel{=}\text{S}}. Since \ensuremath{\mathsf{to}\;({\textbf {\textit {\textrm a}}},{\textbf {\textit {\textrm s}}})\mathrel{=}({\textbf {\textit {\textrm b}}},{\textbf {\textit {\textrm r}}})}, the \ensuremath{{\textbf {\textit {\textrm r}}}\mathbin{:}\overrightarrow{\mathsf{T}}(\Conid{A},\Conid{B})} will actually be
equivalent to \ensuremath{ \delta {\textbf {\textit {\textrm r}}}\mathbin{:}\Varid{a}}, which from~\eqref{eq:incidence} is
equivalent to \ensuremath{\rho {\textbf {\textit {\textrm a}}}\mathbin{:}\Conid{A}}, the post-state of the source update that
is precisely one of the inputs of \ensuremath{\mathsf{to}}.
In such cases, the
transformations will be typed just as follows:
\begin{equation}
  \label{eq:short:scheme}
  \ensuremath{\mathsf{to}\mathbin{:}\overrightarrow{\mathsf{U}}(\Conid{A})\;\!\!\times\!\!\;\overleftarrow{\mathsf{T}}(\Conid{A},\Conid{B})\to \overleftarrow{\mathsf{U}}(\Conid{B})\quad\quad\mathsf{from}\mathbin{:}\overleftarrow{\mathsf{U}}(\Conid{B})\;\!\!\times\!\!\;\overrightarrow{\mathsf{T}}(\Conid{A},\Conid{B})\to \overrightarrow{\mathsf{U}}(\Conid{A})\quad}
\end{equation}

\begin{table}[!t]
\begin{center}
\begin{tabular}{|l|c|p{7cm}|}
    \hline
    \parbox[t]{4.5cm}{\textbf{None} \\\ensuremath{\overrightarrow{\mathsf{T}}(\Conid{A},\Conid{B})\mathrel{=}\mathrm{1}}} & N & No trace information is represented. \\ 
    \hline    
    \parbox[t]{4.5cm}{\textbf{State} \\ \ensuremath{\overrightarrow{\mathsf{T}}(\Conid{A},\Conid{B})\mathrel{=}\Conid{A}}} & S & Only the source state is represented. \\
    \hline
    \parbox[t]{4.5cm}{\textbf{Complement} \\ \ensuremath{\overrightarrow{\mathsf{T}}(\Conid{A},\Conid{B})\mathrel{=}\mathsf{C}(\Conid{A},\Conid{B})}} & C & Some complement of the source and/or target values is represented. \\
    \hline
    \parbox[t]{4.5cm}{\textbf{Delta} \\ \ensuremath{\overrightarrow{\mathsf{T}}(\Conid{A},\Conid{B})\mathrel{=}\Conid{A}\;\!\!\times\!\!\;\Conid{B}\;\!\!\times\!\!\;\mathsf{D}(\Conid{A},\Conid{B})}} & D & Beside both consistent states, a sameness relation between them is represented. \\
    \hline
\end{tabular}
\end{center}
\caption{Traceability.}
\label{tab:traceability}
\vspace{-.2in}
\end{table}

\subsection{Exploring the design space}

By instantiating these two axes we get different flavors of BX
frameworks. As we will present next, some instantiations correspond to
well-known existing frameworks, but others have not been instantiated
yet, and an interesting question is whether they might be useful or
not. 

Not surprisingly, one of the most popular schemes is that of
bidirectional \emph{mappings}, where no traceability information is
represented and updates are typically represented by only the
post-state. Formally, we have \ensuremath{\overrightarrow{\mathsf{U}}\mathrel{=}\overleftarrow{\mathsf{U}}\mathrel{=}\text{S}} and \ensuremath{\overrightarrow{\mathsf{T}}\mathrel{=}\overleftarrow{\mathsf{T}}\mathrel{=}\text{N}}, leading
to the scheme \ensuremath{\mathsf{to}\mathbin{:}\Conid{A}\to \Conid{B}} and \ensuremath{\mathsf{from}\mathbin{:}\Conid{B}\to \Conid{A}}. In these frameworks,
the consistency relation typically corresponds to one of the
transformations and thus is omitted. A symmetric instantiation of this category is that of
bijective languages,
whose transformations establish a bijection between subsets of \ensuremath{\Conid{A}} and
\ensuremath{\Conid{B}}. These subsets contain essentially the same information but just
present it differently. Such languages promote the interoperability
between different formats and are easy to reason about because
bijectivity is preserved by composition and inversion. Less
restrictive asymmetric mapping frameworks encompass transformations
that are only reversible in a particular direction, for example when
\ensuremath{\Conid{B}} refines \ensuremath{\Conid{A}}.

The most popular asymmetric scheme, \emph{lenses}~\cite{Foster:07}, 
was proposed as a solution to the classical view-update problem from database theory~\cite{Bancilhon:81}. 
Updates are still represented using just the post-state, but \ensuremath{\mathsf{from}}
requires traceability information to deal with missing information,
namely the original source value of type \ensuremath{\Conid{A}}. 
More precisely, we have \ensuremath{\overrightarrow{\mathsf{T}}\mathrel{=}\text{S}}
and \ensuremath{\overleftarrow{\mathsf{T}}\mathrel{=}\text{N}}. Since the returned \ensuremath{\overleftarrow{\mathsf{T}}(\Conid{A},\Conid{B})} is equal to the input of \ensuremath{\mathsf{to}}, we
end up with the simplified scheme~\eqref{eq:short:scheme}, resulting in
the interface \ensuremath{\mathsf{to}\mathbin{:}\Conid{A}\to \Conid{B}} and \ensuremath{\mathsf{from}\mathbin{:}\Conid{B}\;\!\!\times\!\!\;\Conid{A}\to \Conid{A}} (known, respectively, as \ensuremath{\Varid{get}}
and \ensuremath{\Varid{put}} in the lens framework).
The consistency relation is assumed to be \ensuremath{(\Varid{a},\Varid{b})\;\!\in\! \;\mathsf{R}} iff \ensuremath{\mathsf{to}\;\Varid{a}\mathrel{=}\Varid{b}}.
In order to guarantee stronger properties or support particular
transformation scenarios, some 
lens-like  
approaches are operation-based 
, or use an additional sameness relation, providing a traceability
between views and sources.

The asymmetric treatment of lenses 
only works well for (essentially) surjective (information decreasing)
transformations, since \ensuremath{\mathsf{to}} cannot access \ensuremath{\Conid{B}} details without
counterpart in \ensuremath{\Conid{A}}.  For more general transformations without a
dominant flow of information, where each of the source and target
models may contain information not present in the other, we end up
with symmetric schemes.  Among those, \emph{maintainers}~\cite{Meertens:98} are among the
most popular, where \ensuremath{\overrightarrow{\mathsf{U}}\mathrel{=}\overleftarrow{\mathsf{U}}\mathrel{=}\text{S}} 
and \ensuremath{\overrightarrow{\mathsf{T}}\mathrel{=}\overleftarrow{\mathsf{T}}\mathrel{=}\text{S}}, leading
to the scheme \ensuremath{\mathsf{to}\mathbin{:}\Conid{A}\;\!\!\times\!\!\;\Conid{B}\to \Conid{B}} and \ensuremath{\mathsf{from}\mathbin{:}\Conid{B}\;\!\!\times\!\!\;\Conid{A}\to \Conid{A}}, where
updates are propagated given knowledge of the pre-state of the
respective opposite transformations. Likewise to
lenses, since the output traceability can trivially be derived from input
updates, it is not returned by the transformations. Also, unlike the
previous frameworks, there is now an explicitly declared consistency relation
\ensuremath{\mathsf{R}}, from which \ensuremath{\mathsf{to}} and \ensuremath{\mathsf{from}} are somehow inferred.

\emph{Trigonal systems} were proposed in~\cite{Diskin:08} to avoid the
recalculation of the whole state values when updates are
incremental. As a generalization of maintainers, besides having
knowledge of the target pre-state, information about the source
pre-state is also present. Concretely, we now have \ensuremath{\overrightarrow{\mathsf{U}}\mathrel{=}\overleftarrow{\mathsf{U}}\mathrel{=}\mathbb{S}} and
\ensuremath{\overrightarrow{\mathsf{T}}\mathrel{=}\overleftarrow{\mathsf{T}}\mathrel{=}\text{S}}. The existence of an explicit
consistency relation \ensuremath{\mathsf{R}} is also assumed.  Like the previous schemes, the content
captured by \ensuremath{\overrightarrow{\mathsf{T}}} is trivially derived from the input information, so
it is not returned by \ensuremath{\mathsf{to}}. However, since the value of the pre-state
of \ensuremath{\overleftarrow{\mathsf{U}}} can also be directly retrieved from \ensuremath{\overleftarrow{\mathsf{T}}}, it is also omitted
from the output.  Putting it all together, we have the scheme \ensuremath{\mathsf{to}\mathbin{:}(\Conid{A}\;\!\!\times\!\!\;\Conid{A})\;\!\!\times\!\!\;\Conid{B}\to \Conid{B}} and \ensuremath{\mathsf{from}\mathbin{:}(\Conid{B}\;\!\!\times\!\!\;\Conid{B})\;\!\!\times\!\!\;\Conid{A}\to \Conid{A}}. 

\emph{Symmetric lenses}~\cite{Hofmann:2011} assume \ensuremath{\overleftarrow{\mathsf{U}}\mathrel{=}\overleftarrow{\mathsf{T}}\mathrel{=}\text{S}} and represent the traceability as a complement (\ensuremath{\overrightarrow{\mathsf{T}}\mathrel{=}\overleftarrow{\mathsf{T}}\mathrel{=}\text{C}}). Thus, we have \ensuremath{\mathsf{to}\mathbin{:}\Conid{A}\;\!\!\times\!\!\;\mathsf{C}(\Conid{A},\Conid{B})\to \Conid{B}\;\!\!\times\!\!\;\mathsf{C}(\Conid{A},\Conid{B})} (and vice-versa), and the complement \ensuremath{\mathsf{C}(\text{S},\Conid{T})} stores both the information of \ensuremath{\Conid{A}} not present in \ensuremath{\Conid{B}} (passed as input to \ensuremath{\mathsf{from}}) and vice-versa.
\emph{Edit lenses}~\cite{Hofmann:12} are an operation-based formulation of symmetric lenses, with edit-sequences as updates (\ensuremath{\overrightarrow{\mathsf{U}}\mathrel{=}\overleftarrow{\mathsf{U}}\mathrel{=}\text{E}}).
We have \ensuremath{\mathsf{to}\mathbin{:}\mathsf{O}(\Conid{A})^\star\;\!\!\times\!\!\;\mathsf{C}(\Conid{A},\Conid{B})\to \mathsf{O}(\Conid{B})^\star\;\!\!\times\!\!\;\mathsf{C}(\Conid{A},\Conid{B})} and the opposite for \ensuremath{\mathsf{from}}.
This time, the transformations do not process states and the complement \ensuremath{\mathsf{C}(\Conid{A},\Conid{B})} stores only some extra information about \ensuremath{\Conid{A}} and \ensuremath{\Conid{B}} that is used to disambiguate updates, but not sufficient to restore the original states.

Among symmetric schemes, \emph{symmetric
  delta-lenses}~\cite{Diskin:11a} are one of the most general. Here
\ensuremath{\overrightarrow{\mathsf{U}}\mathrel{=}\overleftarrow{\mathsf{U}}\mathrel{=}\text{D}} and \ensuremath{\overrightarrow{\mathsf{T}}\mathrel{=}\overleftarrow{\mathsf{T}}\mathrel{=}\text{D}}: besides the pre-state values of \ensuremath{\Conid{A}}
and \ensuremath{\Conid{B}}, we have a sameness relation between them in the traceability,
and likewise for the update itself. We have \ensuremath{\mathsf{to}\mathbin{:}(\Conid{A}\;\!\!\times\!\!\;\Conid{A}\;\!\!\times\!\!\;\mathsf{D}(\Conid{A},\Conid{A}))\;\!\!\times\!\!\;(\Conid{A}\;\!\!\times\!\!\;\Conid{B}\;\!\!\times\!\!\;\mathsf{D}(\Conid{A},\Conid{B}))\to (\Conid{B}\;\!\!\times\!\!\;\Conid{B}\;\!\!\times\!\!\;\mathsf{D}(\Conid{B},\Conid{B}))\;\!\!\times\!\!\;(\Conid{A}\;\!\!\times\!\!\;\Conid{B}\;\!\!\times\!\!\;\mathsf{D}(\Conid{A},\Conid{B}))} and the opposite for \ensuremath{\mathsf{from}}. An
explicit consistency relation \ensuremath{\mathsf{R}} between updates is also present.

\section{Properties}
\label{sec:properties}

The interface of a framework gives some hints about its expressivity
but says little about the actual behavior of the transformations. Such
behavior is usually specified by high-level algebraic properties that
enforce some predictability on the system (namely concerning
bidirectionality). Table~\ref{tab:properties} identifies several generic
properties that, independently of the framework, might be desirable
from an end-user perspective. We only present the properties from the
perspective of \ensuremath{\mathsf{from}} (i.e., propagating updates from the \ensuremath{\Conid{B}} side to
the \ensuremath{\Conid{A}} side). The dual properties can also be specified for \ensuremath{\mathsf{to}}. All
free variables are implicitly universally quantified. This
formalization is to some extent 
a textual version of the graphical
\emph{tile algebra}~\cite{DiskinGTTSE:11}, previously used to
formalize some of the laws presented here. Since the transformation
can be partial, the properties are only required to hold when they yield
a result. Given a transformation $f$, $f\ x \downarrow$ holds when $f$
is defined on $x$, and $f\ x \sqsubseteq y$ holds if $f\ x
\downarrow\ \Rightarrow f\ x = y$.

\begin{table}[!t]
\begin{center}
  \begin{align*}
    & \frac{}{\ensuremath{\mathsf{from}\;({\mathsf{id}_\Conid{B}},{\textbf {\textit {\textrm r}}})\sqsubseteq({\mathsf{id}_\Conid{A}},{{{\textbf {\textit {\textrm r}}}}^\circ})}} \tag{\ensuremath{\mathsf{from}}-Stability}
    \\
    & \frac{\ensuremath{\mathsf{from}\;({\textbf {\textit {\textrm b}}},{\textbf {\textit {\textrm r}}})\mathrel{=}({\textbf {\textit {\textrm a}}},{\textbf {\textit {\textrm s}}})}}{\ensuremath{\mathsf{to}\;({\textbf {\textit {\textrm a}}},{{{\textbf {\textit {\textrm r}}}}^\circ})\sqsubseteq({\textbf {\textit {\textrm b}}},{{{\textbf {\textit {\textrm s}}}}^\circ})}}\tag{\ensuremath{\mathsf{from}}-Invertibility}
    \\
    & \frac{\ensuremath{\mathsf{from}\;({\textbf {\textit {\textrm b}}},{\textbf {\textit {\textrm r}}})\mathrel{=}({\textbf {\textit {\textrm a}}},{\textbf {\textit {\textrm s}}})}}{\ensuremath{\mathsf{from}\;({{{\textbf {\textit {\textrm b}}}}^\circ},{{{\textbf {\textit {\textrm s}}}}^\circ})\sqsubseteq({{{\textbf {\textit {\textrm a}}}}^\circ},{{{\textbf {\textit {\textrm r}}}}^\circ})}}\tag{\ensuremath{\mathsf{from}}-Undoability}   
    \\
    & \frac{\ensuremath{\mathsf{from}\;({\textbf {\textit {\textrm b}}}_{1},{\textbf {\textit {\textrm r}}})\mathrel{=}({\textbf {\textit {\textrm a}}}_{1},{\textbf {\textit {\textrm s}}}_{1})} \quad \ensuremath{\mathsf{from}\;({\textbf {\textit {\textrm b}}}_{2},{{{\textbf {\textit {\textrm s}}}_{1}}^\circ})\mathrel{=}({\textbf {\textit {\textrm a}}}_{2},{\textbf {\textit {\textrm s}}}_{2})}}{\ensuremath{\mathsf{from}\;({\textbf {\textit {\textrm b}}}_{2}\mathbin{\circ}{\textbf {\textit {\textrm b}}}_{1},{\textbf {\textit {\textrm r}}})\sqsubseteq({\textbf {\textit {\textrm a}}}_{2}\mathbin{\circ}{\textbf {\textit {\textrm a}}}_{1},{\textbf {\textit {\textrm s}}}_{2})}}\tag{\ensuremath{\mathsf{from}}-History-ignorance}
    \\
    & \frac{\ensuremath{\mathsf{from}\;({\textbf {\textit {\textrm b}}},{\textbf {\textit {\textrm r}}})\mathrel{=}({\textbf {\textit {\textrm a}}},{\textbf {\textit {\textrm s}}})}}{\ensuremath{{\textbf {\textit {\textrm s}}}\;\!\in\! \;\mathsf{R}}}\tag{\ensuremath{\mathsf{from}}-Correctness}
    \\
    & \frac{\ensuremath{{\textbf {\textit {\textrm b}}}\mathbin{\circ}{\textbf {\textit {\textrm r}}}\;\!\in\! \;\mathsf{R}}}{\ensuremath{\mathsf{from}\;({\textbf {\textit {\textrm b}}},{\textbf {\textit {\textrm r}}})\sqsubseteq({\mathsf{id}_\Conid{A}},{{({\textbf {\textit {\textrm b}}}\mathbin{\circ}{\textbf {\textit {\textrm r}}})}^\circ})}}\tag{\ensuremath{\mathsf{from}}-Hippocraticness}     
    \\
    & \frac{\ensuremath{\mathsf{from}\;({\textbf {\textit {\textrm b}}},{\textbf {\textit {\textrm r}}})\mathrel{=}({\textbf {\textit {\textrm a}}},{\textbf {\textit {\textrm s}}})} \quad \ensuremath{{{{\textbf {\textit {\textrm s}}}}^\circ}\mathbin{\circ}{\textbf {\textit {\textrm a}}}\mathrel{=}{{{\textbf {\textit {\textrm s}}}_{1}}^\circ}\mathbin{\circ}{\textbf {\textit {\textrm a}}}_{1}} \quad \ensuremath{{\textbf {\textit {\textrm s}}}_{1}\;\!\in\! \;\mathsf{R}}}{\ensuremath{{\textbf {\textit {\textrm a}}} \leq {\textbf {\textit {\textrm a}}}_{1}}}\tag{\ensuremath{\mathsf{from}}-Least-update}    
    \\
    & \frac{}{\ensuremath{\mathsf{from}\;({\textbf {\textit {\textrm b}}},{\textbf {\textit {\textrm r}}})} \downarrow}\tag{\ensuremath{\mathsf{from}}-Totality}    
  \end{align*}
\end{center}
\caption{Properties.}
\label{tab:properties}
\vspace{-.2in}
\end{table}

\emph{Stability} imposes that null updates must be translated to
null updates, in the sense that if a \ensuremath{\Conid{B}} is not modified, then no
change shall be performed on the consistent \ensuremath{\Conid{A}}. We represent a null
update on \ensuremath{\Conid{A}} by a constant \ensuremath{{\mathsf{id}_\Conid{A}}}, where \ensuremath{ \delta {\mathsf{id}_\Conid{A}}\mathrel{=}\rho {\mathsf{id}_\Conid{A}}}.
\emph{Invertibility} states that it shall be possible to
revert the application of a transformation by applying the opposite
transformation. This law
implies that updates
on the \ensuremath{\Conid{B}} side are translated faithfully to \ensuremath{\Conid{A}}, otherwise they could
not be inverted. In asymmetric frameworks \ensuremath{\overrightarrow{\mathsf{T}}(\Conid{A},\Conid{B})} might be different
from \ensuremath{\overleftarrow{\mathsf{T}}(\Conid{A},\Conid{B})}. In those cases, we assume that traceability \ensuremath{{\textbf {\textit {\textrm r}}}\mathbin{:}\overrightarrow{\mathsf{T}}(\Conid{A},\Conid{B})} can be reversed as \ensuremath{{{{\textbf {\textit {\textrm r}}}}^\circ}\mathbin{:}\overleftarrow{\mathsf{T}}(\Conid{A},\Conid{B})}, where \ensuremath{ \delta {{{\textbf {\textit {\textrm s}}}}^\circ}\mathrel{=}\rho {\textbf {\textit {\textrm s}}}} and \ensuremath{\rho {{{\textbf {\textit {\textrm s}}}}^\circ}\mathrel{=} \delta {\textbf {\textit {\textrm s}}}}.  \emph{Undoability} ensures that an update
translation can be undone by re-applying the same transformation with
an inverse update. Likewise to traceability, given an update \ensuremath{{\textbf {\textit {\textrm a}}}\mathbin{:}\overrightarrow{\mathsf{U}}(\Conid{A})} its inverse will be denoted by \ensuremath{{{{\textbf {\textit {\textrm a}}}}^\circ}\mathbin{:}\overrightarrow{\mathsf{U}}(\Conid{A})}, where \ensuremath{ \delta {{{\textbf {\textit {\textrm a}}}}^\circ}\mathrel{=}\rho {\textbf {\textit {\textrm a}}}} and \ensuremath{\rho {{{\textbf {\textit {\textrm a}}}}^\circ}\mathrel{=} \delta {\textbf {\textit {\textrm a}}}}. Also, two
updates \ensuremath{{\textbf {\textit {\textrm a}}}_{1}\mathbin{:}\overrightarrow{\mathsf{U}}(\Conid{A})} and \ensuremath{{\textbf {\textit {\textrm a}}}_{2}\mathbin{:}\overrightarrow{\mathsf{U}}(\Conid{A})} can be sequentially composed
as \ensuremath{{\textbf {\textit {\textrm a}}}_{2}\mathbin{\circ}{\textbf {\textit {\textrm a}}}_{1}\mathbin{:}\overrightarrow{\mathsf{U}}(\Conid{A})}.
\emph{History-ignorance} states that update
translation does not depend on the past history. In practice, this
means that two consecutive update translations can be performed at
once on the composed update.

The following three properties involve the
consistency relation. \emph{Correctness} simply states that a
transformation restores consistency. \emph{Hippocraticness} is a
stronger version of stability (although not always
desirable~\cite{Diskin:08}), stating that an update that does not
break the consistency should be ignored. In the generic formulation we
assume that a traceability \ensuremath{{\textbf {\textit {\textrm r}}}\mathbin{:}\overrightarrow{\mathsf{T}}(\Conid{A},\Conid{B})} can be composed with an
update \ensuremath{{\textbf {\textit {\textrm b}}}\mathbin{:}\overrightarrow{\mathsf{U}}(\Conid{B})} to yield a traceability \ensuremath{{\textbf {\textit {\textrm b}}}\mathbin{\circ}{\textbf {\textit {\textrm r}}}\mathbin{:}\overrightarrow{\mathsf{T}}(\Conid{A},\Conid{B})}
relating the original source \ensuremath{\Conid{A}} with the updated \ensuremath{\Conid{B}}. Due to the
incidence conditions, checking the consistency of \ensuremath{{\textbf {\textit {\textrm b}}}\mathbin{\circ}{\textbf {\textit {\textrm r}}}} is
equivalent to checking the consistency of these values.  
In schemes where the consistency relation is one of the transformations,
correctness and hippocraticness degenerate into invertibility and stability, respectively.
The \emph{least-update} property can be seen as an additional \emph{quality property} entailing that the returned update must be
the smallest among all \ensuremath{\mathsf{R}}-consistent ones that could have been returned.
To compare updates, we assume the existence of a total
preorder \ensuremath{ \leq } on \ensuremath{\overrightarrow{\mathsf{U}}(\Conid{A})}. When \ensuremath{\mathsf{from}\;({\textbf {\textit {\textrm b}}},{\textbf {\textit {\textrm r}}})\mathrel{=}({\textbf {\textit {\textrm a}}},{\textbf {\textit {\textrm s}}})} then
\ensuremath{{\textbf {\textit {\textrm a}}}} must be smaller than every update \ensuremath{{\textbf {\textit {\textrm a}}}_{1}} that could lead to a
value consistent with the post-state of \ensuremath{{\textbf {\textit {\textrm b}}}} (testified by a
consistent traceability \ensuremath{{\textbf {\textit {\textrm s}}}_{1}}).  Assuming that, for an already consistent state, the null target update
is the unique minimal update, least-update subsumes hippocraticness.

So far, we presented the properties modulo undefinedness of the
unidirectional transformations.  This is because totality requirements
are by themselves an important feature in the design of a BX
framework.  In practice, it is often convenient to acknowledge that
the type system might not be expressive enough to capture all
constraints induced by the transformations, and to allow the source
and target types to be larger than the actual domains of the
transformations, leading to partially defined transformations.  
While this might be a ``show stopper'' for batch applications that are
expected to always produce results, it is usually acceptable for
interactive applications: an editor does not need to handle every
update and can signal an error to the user disallowing a specific
modification.  Partiality is not adequate for security
applications~\cite{Foster:2009b} though, since users might extract
information about the hidden data from the cases for which the
transformations fail.  As such, to allow a finer-grain comparison of
frameworks, we choose to factor out \emph{totality} as an orthogonal
property: it holds for a transformation if it is defined for every
possible combination of update and traceability.

Sometimes, weaker versions of the above laws may be satisfied instead. For
example, we can have weaker versions of \emph{invertibility} where the
final state is equal to the original one modulo another update
translation. This particular \emph{weak invertibility} is a kind of
\emph{convergence} law (or \emph{bi-idempotence}
in~\cite{Hu:04journal}), since it entails that update translation
eventually converges into stable states.  Other \emph{weak} variants
of the laws occur when value comparison ignores details
that are inessential for an application scenario, like ordering, whitespaces or structure sharing.
For operation-based frameworks, these
may also mean that round-tripping does not preserve the full update,
but only its post-state. 
An interesting weaker
version of totality is \emph{safety}~\cite{Pacheco:12} (also known as \emph{domain correctness}
in~\cite{Diskin:08}), which entails that a transformation is defined
at least for the range of the opposite one, independently of its
pre-state. Taking into account the consistency relation, safety can also be stated as follows: a transformation must be defined for every source value that has at least one consistent target.
Weaker versions of \emph{correctness} include allowing the creation of inconsistent states if no consistent ones exists.

\subsection{Revisiting the design space}

To instantiate the generic properties for a concrete framework, one
must first devise how to express null, inversion and composition of
updates and traceability. In some frameworks some of these might not
be expressible, meaning that some properties may not be applicable.
For example, \emph{mappings} have no traceability and updates are
represented just by the post-state, hence there is no way to reason
about pre-states. This implies that neither null updates, nor update
inversion can be defined, and \ensuremath{\mathsf{from}}-Stability, \ensuremath{\mathsf{from}}-Undoability,
and \ensuremath{\mathsf{from}}-Hippocractiness are not expressible.  \ensuremath{\mathsf{from}}-Invertibility
is just \ensuremath{\mathsf{to}\;(\mathsf{from}\;\Varid{b})\sqsubseteq\Varid{b}}, ensuring \ensuremath{\mathsf{from}} to be left-invertible
(injective).  Without any knowledge of the pre-state,
\ensuremath{\mathsf{from}}-History-ignorance holds trivially. Assuming the consistency relation \ensuremath{(\Varid{a},\Varid{b})\;\!\in\! \;\mathsf{R}\equiv (\Varid{b}\mathrel{=}\mathsf{to}\;\Varid{a})}, \ensuremath{\mathsf{from}}-Correctness
degenerates to \ensuremath{\mathsf{from}}-Invertibility, and \ensuremath{\mathsf{from}}-Least-update amounts
to checking that if \ensuremath{\mathsf{from}\;\Varid{b}\mathrel{=}\Varid{a}}, then \ensuremath{\Varid{a}} is smaller than every \ensuremath{\Varid{a}_{1}}
leading to the same \ensuremath{\Varid{b}}, that is \ensuremath{\mathsf{from}\;\Varid{b}\mathrel{=}\Varid{a}\;\wedge\;\mathsf{to}\;\Varid{a}\mathrel{=}\mathsf{to}\;\Varid{a}_{1}\mathrel{=}\Varid{b}\Rightarrow \Varid{a} \leq \Varid{a}_{1}}.

Unlike mappings, in \emph{lenses} there is some traceability
information when applying \ensuremath{\mathsf{from}} that allows us to reason about
pre-states. In particular, when \ensuremath{\mathsf{from}} is applied to traceability \ensuremath{\Varid{a}}
(the original source value), due to the consistency relation
\ensuremath{(\Varid{a},\Varid{b})\;\!\in\! \;\mathsf{R}\equiv (\Varid{b}\mathrel{=}\mathsf{to}\;\Varid{a})}, we know that the pre-state of the input
update is \ensuremath{\mathsf{to}\;\Varid{a}}. Hence, a null update has the same
post-state, and \ensuremath{\mathsf{from}}-Stability can be instantiated as \ensuremath{\mathsf{from}\;(\mathsf{to}\;\Varid{a},\Varid{a})\sqsubseteq\Varid{a}} (known in this framework as \textsc{GetPut}).
Instantiation of \ensuremath{\mathsf{from}}-Invertibility is more
straightforward --- just ignore unused traceability --- yielding \ensuremath{\mathsf{from}\;(\Varid{b}_{1},\anonymous )\mathrel{=}\Varid{a}_{1}\Rightarrow \mathsf{to}\;\Varid{a}_{1}\sqsubseteq\Varid{b}_{1}} (known as \textsc{PutGet}). 
To instantiate \ensuremath{\mathsf{from}}-Undoability, we follow the same approach as in \ensuremath{\mathsf{from}}-Stability, resulting in \ensuremath{\mathsf{from}\;(\Varid{b}_{1},\Varid{a})\mathrel{=}\Varid{a}_{1}\Rightarrow \mathsf{from}\;(\mathsf{to}\;\Varid{a},\Varid{a}_{1})\sqsubseteq\Varid{a}}.
Since the composition of two state-based updates \ensuremath{\Varid{b}\mathbin{\circ}\Varid{a}} is just \ensuremath{\Varid{b}},
instantiation of \ensuremath{\mathsf{from}}-History-Ignorance is \ensuremath{\mathsf{from}\;(\Varid{b}_{1},\Varid{a})\mathrel{=}\Varid{a}_{1}\mathrel{\wedge}\mathsf{from}\;(\Varid{b}_{2},\Varid{a}_{1})\mathrel{=}\Varid{a}_{2}\Rightarrow \mathsf{from}\;(\Varid{b}_{2},\Varid{a})\sqsubseteq\Varid{a}_{2}} (known as
\textsc{PutPut}).  Due to the consistency
relation being \ensuremath{\mathsf{to}}, \ensuremath{\mathsf{from}}-Correctness and \ensuremath{\mathsf{from}}-Hippocraticness degenerate
into \ensuremath{\mathsf{from}}-Invertibility and \ensuremath{\mathsf{from}}-Stability, respectively. Likewise
to mappings, \ensuremath{\mathsf{from}}-Least-update is formulated as \ensuremath{\mathsf{from}\;(\Varid{b}_{1},\Varid{a})\mathrel{=}\Varid{a}_{1}\;\wedge\;\mathsf{to}\;\Varid{a}_{2}\mathrel{=}\Varid{b}_{1}\Rightarrow \Varid{a}_{1} \leq \Varid{a}_{2}}.

Like in the previous frameworks, the value of the pre-state of the
input update is not explicitly represented in \emph{maintainers}.
However, unlike lenses, the declared consistency relation of a maintainer may
not be functional (deterministic), and a value of \ensuremath{\Conid{A}} is not uniquely related to another value
of \ensuremath{\Conid{B}}. This means that it is impossible to identify a null update given
only the original source value present in the traceability, and
\ensuremath{\mathsf{from}}-Stability cannot be formulated.
In the case of \ensuremath{\mathsf{from}}-Invertibility, to revert a transformation \ensuremath{\mathsf{from}\;(\Varid{b}_{1},\Varid{a})\mathrel{=}\Varid{a}_{1}} we need to invert the traceability \ensuremath{\Varid{a}}, and recover the
original consistent \ensuremath{\Varid{b}}. As discussed above, in general this is not
possible, but we can generalize this property assuming that the
transformation can be reverted for any consistent \ensuremath{\Varid{b}}, that is \ensuremath{(\Varid{a},\Varid{b})\;\!\in\! \;\mathsf{R}\;\wedge\;\mathsf{from}\;(\Varid{b}_{1},\Varid{a})\mathrel{=}\Varid{a}_{1}\Rightarrow \mathsf{to}\;(\Varid{a}_{1},\Varid{b})\sqsubseteq\Varid{b}_{1}}.
Following a similar approach, \ensuremath{\mathsf{from}}-Undoability can be formulated as \ensuremath{(\Varid{a},\Varid{b})\;\!\in\! \;\mathsf{R}\;\wedge\;\mathsf{from}\;(\Varid{b}_{1},\Varid{a})\sqsubseteq\Varid{a}_{1}\Rightarrow \mathsf{from}\;(\Varid{b},\Varid{a}_{1})\sqsubseteq\Varid{a}}. 
Since the \ensuremath{\mathsf{from}} interface is similar to lenses, \ensuremath{\mathsf{from}}-History-ignorance is exactly the same.
Due to the explicit consistency relation,
\ensuremath{\mathsf{from}}-Correctness is directly formulated as \ensuremath{\mathsf{from}\;(\Varid{b},\anonymous )\mathrel{=}\Varid{a}\Rightarrow (\Varid{a},\Varid{b})\;\!\in\! \;\mathsf{R}} and \ensuremath{\mathsf{from}}-Hippocraticness as \ensuremath{(\Varid{a},\Varid{b})\;\!\in\! \;\mathsf{R}\Rightarrow \mathsf{from}\;(\Varid{b},\Varid{a})\sqsubseteq\Varid{a}}. Lastly, \ensuremath{\mathsf{from}}-Least-update is again similar to that of lenses but with an explicit consistency relation, i.e., \ensuremath{\mathsf{from}\;(\Varid{b},\Varid{a})\mathrel{=}\Varid{a}_{1}\;\wedge\;(\Varid{a}_{2},\Varid{b})\;\!\in\! \;\mathsf{R}\Rightarrow \Varid{a}_{1} \leq \Varid{a}_{2}}.

In \emph{trigonal systems}, updates are represented by both pre- and
post-state value. As such, null updates and composition and inversion
of updates can be directly defined as \ensuremath{{\mathsf{id}_\Conid{B}}\mathrel{=}(\Varid{b},\Varid{b})},
\ensuremath{(\Varid{b}_{1},\Varid{b}_{2})\mathbin{\circ}(\Varid{b}_{2},\Varid{b}_{3})\mathrel{=}(\Varid{b}_{1},\Varid{b}_{3})} and \ensuremath{{{(\Varid{b}_{1},\Varid{b}_{2})}^\circ}\mathrel{=}(\Varid{b}_{2},\Varid{b}_{1})}, and the
instantiation of properties becomes rather straightforward. Namely,
\ensuremath{\mathsf{from}}-stability is \ensuremath{\mathsf{from}\;((\Varid{b},\Varid{b}),\Varid{a})\sqsubseteq\Varid{a}}, \ensuremath{\mathsf{from}}-Invertibility is
\ensuremath{\mathsf{from}\;((\Varid{b},\Varid{b}_{1}),\Varid{a})\mathrel{=}\Varid{a}_{1}\Rightarrow \mathsf{to}\;((\Varid{a},\Varid{a}_{1}),\Varid{b})\sqsubseteq\Varid{b}_{1}}, \ensuremath{\mathsf{from}}-Undoability
is \ensuremath{\mathsf{from}\;((\Varid{b},\Varid{b}_{1}),\Varid{a})\mathrel{=}\Varid{a}_{1}\Rightarrow \mathsf{from}\;((\Varid{b}_{1},\Varid{b}),\Varid{a}_{1})\sqsubseteq\Varid{a}}, and
\ensuremath{\mathsf{from}}-History-ignorance is \ensuremath{\mathsf{from}\;((\Varid{b},\Varid{b}_{1}),\Varid{a})\mathrel{=}\Varid{a}_{1}\;\wedge\;\mathsf{from}\;((\Varid{b}_{1},\Varid{b}_{2}),\Varid{a}_{1})\mathrel{=}\Varid{a}_{2}\Rightarrow \mathsf{from}\;((\Varid{b},\Varid{b}_{2}),\Varid{a})\sqsubseteq\Varid{a}_{2}}. Likewise to
maintainers, due to the explicit consistency relation, the remaining
instantiations are also immediate: \ensuremath{\mathsf{from}}-Correctness is \ensuremath{\mathsf{from}\;((\anonymous ,\Varid{b}_{1}),\anonymous )\mathrel{=}\Varid{a}_{1}\Rightarrow (\Varid{a}_{1},\Varid{b}_{1})\;\!\in\! \;\mathsf{R}}, \ensuremath{\mathsf{from}}-Hippocracticness is \ensuremath{(\Varid{a},\Varid{b})\;\!\in\! \;\mathsf{R}\Rightarrow \mathsf{from}\;((\anonymous ,\Varid{b}_{1}),\Varid{a})\sqsubseteq\Varid{a}} and \ensuremath{\mathsf{from}}-Least-update is \ensuremath{\mathsf{from}\;((\anonymous ,\Varid{b}_{1}),\Varid{a})\mathrel{=}\Varid{a}_{1}\;\wedge\;(\Varid{a}_{2},\Varid{b}_{1})\;\!\in\! \;\mathsf{R}\Rightarrow (\Varid{a},\Varid{a}_{1}) \leq (\Varid{a},\Varid{a}_{2})}. Likewise, the instantiation
of the properties in the framework of \emph{symmetric delta-lenses} is
straightforward, since both the pre- and post-state values of
updates are represented and have an explicit consistency relation.

In \emph{edit lenses}, updates are represented by sequences of edit
operations: the null update is the empty sequence \ensuremath{[\mskip1.5mu \mskip1.5mu]}, composing
updates \ensuremath{{\textbf {\textit {\textrm a}}}} and \ensuremath{{\textbf {\textit {\textrm b}}}} amounts to concatenation \ensuremath{{\textbf {\textit {\textrm a}}}\plus {\textbf {\textit {\textrm b}}}}, and assuming each
edit operation to be undoable, an update \ensuremath{{\textbf {\textit {\textrm a}}}\mathrel{=}[\mskip1.5mu \Varid{a}_{1},\mathinner{\ldotp\ldotp},\Varid{a}_{\Varid{n}}\mskip1.5mu]} could be
inverted, for example, as \ensuremath{{{{\textbf {\textit {\textrm a}}}}^\circ}\mathrel{=}[\mskip1.5mu {{\Varid{a}_{\Varid{n}}}^\circ},\mathinner{\ldotp\ldotp},{{\Varid{a}_{1}}^\circ}\mskip1.5mu]}.
Traceability information (the complement) is ``symmetric'' in the sense that \ensuremath{{{{\textbf {\textit {\textrm c}}}}^\circ}\mathrel{=}{\textbf {\textit {\textrm c}}}}.
Equipped with these definitions, some properties can be
directly instantiated: \ensuremath{\mathsf{from}}-Stability is \ensuremath{\mathsf{from}\;([\mskip1.5mu \mskip1.5mu],{\textbf {\textit {\textrm c}}})\sqsubseteq([\mskip1.5mu \mskip1.5mu],{\textbf {\textit {\textrm c}}})},
\ensuremath{\mathsf{from}}-Invertibility is \ensuremath{\mathsf{from}\;({\textbf {\textit {\textrm b}}}_{1},{\textbf {\textit {\textrm c}}})\mathrel{=}({\textbf {\textit {\textrm a}}}_{1},{\textbf {\textit {\textrm c}}}_{1})\Rightarrow \mathsf{to}\;({\textbf {\textit {\textrm a}}}_{1},{\textbf {\textit {\textrm c}}})\sqsubseteq({\textbf {\textit {\textrm b}}}_{1},{\textbf {\textit {\textrm c}}}_{1})},
\ensuremath{\mathsf{from}}-Undoability is \ensuremath{\mathsf{from}\;({\textbf {\textit {\textrm b}}}_{1},{\textbf {\textit {\textrm c}}})\mathrel{=}({\textbf {\textit {\textrm a}}}_{1},{\textbf {\textit {\textrm c}}}_{1})\Rightarrow \mathsf{from}\;({{{\textbf {\textit {\textrm b}}}_{1}}^\circ},{\textbf {\textit {\textrm c}}}_{1})\sqsubseteq({{{\textbf {\textit {\textrm a}}}_{1}}^\circ},{\textbf {\textit {\textrm c}}})}, and \ensuremath{\mathsf{from}}-History-ignorance is \ensuremath{\mathsf{from}\;({\textbf {\textit {\textrm b}}}_{1},{\textbf {\textit {\textrm c}}})\mathrel{=}({\textbf {\textit {\textrm a}}}_{1},{\textbf {\textit {\textrm c}}}_{1})\;\wedge\;\mathsf{from}\;({\textbf {\textit {\textrm b}}}_{2},{\textbf {\textit {\textrm c}}}_{1})\mathrel{=}({\textbf {\textit {\textrm a}}}_{2},{\textbf {\textit {\textrm c}}}_{2})\Rightarrow \mathsf{from}\;({\textbf {\textit {\textrm b}}}_{1}\plus {\textbf {\textit {\textrm b}}}_{2},{\textbf {\textit {\textrm c}}})\sqsubseteq({\textbf {\textit {\textrm a}}}_{1}\plus {\textbf {\textit {\textrm a}}}_{2},{\textbf {\textit {\textrm c}}}_{2})}. 
In \ensuremath{\mathsf{from}}-Correctness the existence of the initial source-target pair \ensuremath{(\Varid{a},\Varid{b})} that gave origin to complement \ensuremath{{\textbf {\textit {\textrm c}}}} is assumed, over which the resulting edit-sequences are applied, resulting in \ensuremath{(\Varid{a},\Varid{b})\;\!\in\! \;\mathsf{R}\mathrel{\wedge}\mathsf{from}\;({\textbf {\textit {\textrm b}}}_{1},{\textbf {\textit {\textrm c}}})\mathrel{=}({\textbf {\textit {\textrm a}}}_{1},{\textbf {\textit {\textrm c}}}_{1})\Rightarrow ({\textbf {\textit {\textrm a}}}_{1}\;\Varid{a},{\textbf {\textit {\textrm b}}}_{1}\;\Varid{b})\;\!\in\! \;\mathsf{R}}, where \ensuremath{({\textbf {\textit {\textrm a}}}\;\Varid{a})} denotes the application of the edit-sequence \ensuremath{{\textbf {\textit {\textrm a}}}} to the value \ensuremath{\Varid{a}}.
The information about the system state is external to the transformations, as updates are only represented by edit-sequences.
\ensuremath{\mathsf{from}}-Hippocracticness applies to transformations over the complement of states that were already consistent, represented as
\ensuremath{(\Varid{a},\Varid{b})\;\!\in\! \;\mathsf{R}\Rightarrow \mathsf{from}\;({\textbf {\textit {\textrm b}}}_{1},{\textbf {\textit {\textrm c}}})\sqsubseteq([\mskip1.5mu \mskip1.5mu],{\textbf {\textit {\textrm c}}})}. Lastly, \ensuremath{\mathsf{from}}-Least-update is formulated as 
\ensuremath{(\Varid{a},\Varid{b})\;\!\in\! \;\mathsf{R}\mathrel{\wedge}\mathsf{from}\;({\textbf {\textit {\textrm b}}}_{1},{\textbf {\textit {\textrm c}}})\mathrel{=}({\textbf {\textit {\textrm a}}}_{1},{\textbf {\textit {\textrm c}}}_{1})\mathrel{\wedge}({\textbf {\textit {\textrm a}}}_{2}\;\Varid{a},{\textbf {\textit {\textrm b}}}_{1}\;\Varid{b})\;\!\in\! \;\mathsf{R}\Rightarrow {\textbf {\textit {\textrm a}}}_{1} \leq {\textbf {\textit {\textrm a}}}_{2}}. Note that the pre-order compares only edit-sequences.

Each framework has a notion of what is a \emph{well-behaved}
transformation, i.e., the minimum properties it must satisfy to be
considered reasonable. Typically, a transformation is
\emph{well-behaved} if it is at least stable and correct, i.e.,
preserves null updates and recovers consistency. In mappings
\ensuremath{\mathsf{to}}-Correctness degenerates into \ensuremath{\mathsf{to}}-Invertibility, meaning that
\ensuremath{\mathsf{to}} is injective and \ensuremath{\Conid{B}} is a \emph{refinement} of \ensuremath{\Conid{A}} (it
contains more information).  In a \emph{well-behaved} symmetric
mapping, both \ensuremath{\mathsf{to}}-Invertibility and \ensuremath{\mathsf{from}}-Invertibility hold and
the BX is an \emph{isomorphism}, such that \ensuremath{\mathsf{to}} and \ensuremath{\mathsf{from}}
are bijections (when restricted to the respective domains).
In asymmetric lenses, \ensuremath{\mathsf{from}}-Invertibility together with
\ensuremath{\mathsf{from}}-Stability are the typical laws required for a lens to be
well-behaved. 
The law \ensuremath{\mathsf{from}}-Invertibility implies that \ensuremath{\mathsf{to}} is surjective (again when restricted to the respective domain) and \ensuremath{\Conid{B}} is an
\emph{abstraction} of \ensuremath{\Conid{A}} (also called a view), meaning that the target contains less information than the source. If a framework also satisfies
history ignorance, then it is usually considered \emph{very well-behaved}. Since stability and history ignorance entail undoability,
very well-behaved frameworks are also undoable.

Some approaches do not enforce any totality requirements
. However, this can easily be abused: a (partial) BX can be
trivially well-behaved if both transformations are always
undefined.
For asymmetric frameworks, one transformation generally dominates the data flow and has stronger totality requirements than the other, and thus approaches usually assume \ensuremath{\mathsf{to}} to be total and \ensuremath{\mathsf{from}} either partial
or safe.
For symmetric frameworks, there is not generally a dominant data flow (for example, not every Java feature can be represented with a relational database schema, and vice-versa), and both transformations may be plausibly partial.
For \emph{total} BXs, 
the types capture the exact domains over which the transformation is defined and guaranteed to behave well, ensuring that update translation cannot fail at run time.

\section{A survey of existing BX frameworks}
\label{sec:survey-existing-bx}

\begin{table}[!thb]
\begin{center}
\makebox[\textwidth]{%
\begin{tabular}{|c!{\vrule width 1.5pt}c|cc|cc|c||c|c|c|c|c|c|c|c|c|}
	\hline
   \multirow{2}{*}{
	\begin{picture}(130,75)(0,0)
         \put(85,55){\bf Feature}
         \put(-6,75){\line(5,-3){140}}
         \put(10,-10){\bf Approach}
       \end{picture}}
   & \multicolumn{6}{c||}{\hyperref[sec:framework]{Scheme}} & \multicolumn{9}{c|}{\hyperref[sec:properties]{Properties}} \\
\cline{2-16}
   &\raiserot{\hyperref[tab:symmetry]{Symmetry}} & \raiserot{\hyperref[tab:update]{\ensuremath{\overrightarrow{\mathsf{U}}}}} & \raiserot{\hyperref[tab:update]{\ensuremath{\overleftarrow{\mathsf{U}}}}} & \raiserot{\hyperref[tab:traceability]{\ensuremath{\overleftarrow{\mathsf{T}}}}} & \raiserot{\hyperref[tab:traceability]{\ensuremath{\overrightarrow{\mathsf{T}}}}} & \raiserot{\hyperref[tab:consistency]{\ensuremath{\mathsf{R}}}} 
   &\raiserot{\hyperref[tab:properties]{Stable}} & \raiserot{\hyperref[tab:properties]{Invertible}} & \raiserot{\hyperref[tab:properties]{Convergent}} & \raiserot{\hyperref[tab:properties]{Undoable}} 
   &\raiserot{\hyperref[tab:properties]{History Ignorant}} & \raiserot{\hyperref[tab:properties]{Correct}} & \raiserot{\hyperref[tab:properties]{Hippocratic}} & \raiserot{\hyperref[tab:properties]{Least-update}} & \raiserot{\hyperref[tab:properties]{Total}} \\
\hline\hline
Brabrand et al. (2008)~\cite{Brabrand:2008}&		    S&	\mcb{S}&	\mcb{N}&	E&  &			$\dashes$&	&			&			&			$\arrows$&	&			&           $\arrows$	\\ 
Kawanaka and Hosoya (2006)~\cite{Kawanaka:2006} &    S&  \mcb{S}&	\mcb{N}&	E&  &			&			&			&			&			$\arrows$&	&			&           $\arrows$	\\ 
Ehrig et al. (2007)~\cite{Ehrig:2007}&				S&	\mcb{S}&	\mcb{D}&	E&  &			$\arrows$&	&			&			&			$\arrows$&	&			&           $\arrows$	\\ 
Wadler (1987)~\cite{Wadler:1987} &					S& 	\mcb{S}&	\mcb{N}&	T&  &			$\arrows$&	&			&			&			&			&			&           			\\ 
Atanassow and Jeuring (2007)~\cite{Atanassow:2007}&	S&  \mcb{S}&	\mcb{N}&	I&  &			$\arrows$&	&			&			&			&			&			&           			\\ 
Kennedy (2004)~\cite{Kennedy:2004} &				    A&	\mcb{S}&	\mcb{N}&	T&  &			$\ldashes$&	&			&			&			&			&			&           			\\ 
Terwilliger et al. (2007)~\cite{Terwilliger:2007} &	A&  S&	E&		\mcb{N}&	T&  &			$\rarrow$&	&			&			&			&			&			&           $\ldashes$	\\ 
Cunha et al. (2012)~\cite{Cunha:12} &				A&  \mcb{E}&	\mcb{N}&	E&  $\arrows$&	$\rarrow$&	&       	&			$\arrows$&	$\arrows$&	$\larrow$&	&           $\arrows$	\\ 
Mu et al. (2004)~\cite{Mu:04} &				        A&  S&	E&		\mcb{N}&	T&  &			$\rarrow$&	$\ldash$&	&			&			&			&			&           			\\ 
Berdaguer et al. (2007)~\cite{Berdaguer:07} &	    A&  \mcb{S}&	\mcb{N}&	T&  &			$\rarrow$&	&			&			&			&			&			&           $\ldashes$	\\ 
Wang et al. (2010)~\cite{Wang:2010} &			    A&  \mcb{S}&	\mcb{N}&	T&  &			$\larrow$&	&			&			&			&			&			&           $\arrows$	\\ 
Foster et al. (2007)~\cite{Foster:07} &				    A&	\mcb{S}&	N&	S&		T&  $\arrows$&	$\larrow$&  &			&			&			&			&			&           $\arrows$	\\ 
Bohannon et al. (2006)~\cite{Bohannon:06} &			A&	\mcb{S}&	N&	S&		T&  $\arrows$&	$\larrow$&  &			&			&			&			&			&           $\arrows$	\\ 
Bohannon et al. (2008)~\cite{Bohannon:08} &		    A&	\mcb{S}&	N&	S&		T&  $\ldashes$&	$\larrow$&	&			&			&			&			&			&           $\arrows$	\\ 
Foster et al. (2008)~\cite{Foster:2008} &		    A&	\mcb{S}&	N&	S&		T&  $\dashes$&	$\ldash$&   &			&			&			&			&			&           $\arrows$	\\ 
Barbosa et al. (2010)~\cite{Barbosa:2010} &		    A&	S&	D&		N&	D&		T&  $\arrows$&	$\larrow$& 	&			&			&			&			&			&           $\arrows$	\\ 
Hu et al. (2008)~\cite{Hu:04journal} &				    A&	S&	E&		N&	S&		T&  &			&		    $\dashes$&	&			&			&			&			&           $\rarrow$	\\ 
Liu et al. (2007)~\cite{Liu:07} &				    A&	S&	E&		N&	S&		T&  $\arrows$&	$\larrow$&	&			&			&			&			&			&           $\rarrow$	\\ 
Hidaka et al. (2010)~\cite{Hidaka:10} &			    A&	S&	E&		N&	S&		T&  $\larrow$&	$\ldash$&	&			&			&			&			&			&           $\rarrow$	\\ 
Takeichi (2009)~\cite{Takeichi:09} &				    A&	\mcb{S}&	N&	S&		I&  $\larrow$&	&		    &			&			&			&			&			&           $\rarrow$	\\ 
Matsuda et al. (2007)~\cite{Matsuda:07} &		    A&	\mcb{S}&	N&	C&  	T&  $\arrows$&	$\larrow$&	&			&			$\larrow$&	&			&			&           $\rarrow$	\\ 
Voigtl\"{a}nder (2009)~\cite{Voigtlander:09} &	    A&	\mcb{S}&	N&	S&		T&  $\arrows$&	$\larrow$&  &			&			$\larrow$&	&			&			&           $\rarrow$	\\ 
Voigtl\"{a}nder et al. (2010)~\cite{Voigtlander:10}& A&	\mcb{S}&	N&	S&		T&  $\arrows$&	$\larrow$&  &			&			&			&			&			&           $\rarrow$	\\ 
Fegaras (2010)~\cite{Fegaras:10}&				    A&	S&	E&  	N&	D&		T&  $\larrow$&	$\ldash$&   &			&			&			&			&			&           $\rarrow$	\\ 
Melnik et al. (2007)~\cite{Melnik:07} &			    A&	\mcb{S}&	N&	S&		E&  $\arrows$&	$\larrow$& 	&			&			&			&			&			&           $\rdashes$	\\ 
Diskin et al. (2011)~\cite{Diskin:11a}* &			    A&	\mcb{D}&	N&	S&		T&  $\arrows$&	$\larrow$&  &			&			$\arrows$&	&			&			&           $\ldashes$	\\ 
Wang et al. (2011)~\cite{Wang:2011}* &			    A&	S&	F&		N&	S&		T&  $\arrows$&	$\larrow$&  &			$\larrow$&	&			&			&			&           $\rarrow$	\\ 
Pacheco and Cunha (2010)~\cite{PachecoCunha:10} &    A&	\mcb{S}&	N&	S&		T&  $\arrows$&	$\larrow$&  &			&			&			&			&			&           $\arrows$	\\ 
Pacheco and Cunha (2012)~\cite{PachecoCunha:12} &    A&	\mcb{D}&	N&	S&		T&  $\arrows$&	$\larrow$&  &			&			&			&			&			&           $\arrows$	\\ 
Meertens (1998)~\cite{Meertens:98} &				    S&	\mcb{S}&	\mcb{S}&	E&  &       	&			&			&			&			$\arrows$&	$\arrows$&	&           $\arrows$	\\ 
Meertens (1998)~\cite{Meertens:98}&	    			    S&	\mcb{E}&	\mcb{S}&	E&  &       	&			&			&			&			$\arrows$&	&	        $\arrows$&  $\arrows$	\\ 
Stevens (2007)~\cite{Stevens:07}* &				    S&	\mcb{S}&	\mcb{S}&	E&  &	        &			&			$\arrows$&	&			$\arrows$&	$\arrows$&	&           $\arrows$	\\ 
Macedo and Cunha (2013)~\cite{MacedoCunha:13} &		S&	\mcb{S}&	\mcb{S}&	E&  &           &			&			&       	&			$\arrows$&  &           $\arrows$&  $\dashes$  	\\ 
Hofmann et al. (2011)~\cite{Hofmann:2011} &			    S&	\mcb{S}&	\mcb{C}&    I&  $\arrows$&	&			&			&			&			&			&			&           $\arrows$	\\ 
Hofmann et al. (2012)~\cite{Hofmann:12}&			    S&	\mcb{E}&	\mcb{C}&    I&  &			&			&			&			&			$\arrows$&	&			&           $\dashes$	\\ 
Diskin et al. (2011)~\cite{Diskin:2011}* &			    S&	\mcb{D}&	\mcb{D}&	T&  &			&			$\arrows$&	$\dashes$&	&			$\arrows$&	&		 	&           		    \\ 
Hermann et al. (2011)~\cite{Hermann:2011} &		    S&	\mcb{D}&	\mcb{D}&	E&  &			&			$\arrows$&	&			&			$\arrows$&	&			&           $\dashes$	\\ 
Cicchetti et al. (2011)~\cite{Cicchetti:11} &		S&	\mcb{S}&	\mcb{S}&	E&  $\arrows$&	&			&			&			&$\dashes$			&	        &			&           $\arrows$	\\ 
Ennals and Gay (2007)~\cite{Ennals:2007} &		    S&	\mcb{S}&	\mcb{S}&	I&  $\dashes$&	&			&			&			&			$\arrows$&	&			&           $\arrows$	\\ 
\hline
\end{tabular}}
\end{center}
\caption{Comparison of existing BX approaches.}
\label{table:stateofart}
\end{table}

Based on the proposed generic scheme and properties, we attempted a
comprehensive survey of existing BX tools and
frameworks. Table~\ref{table:stateofart} presents the results of this
first effort\footnote{A more detailed classification of BX approaches related to this paper can be found in~\cite[Chapter 3]{Pacheco:12}. Although the mentioned scheme regards only 3 specific frameworks (mappings, lenses and maintainers), this complementary work also proposes a taxonomy for the particular deployment features of BX frameworks and a textual justification for each of the entries in Table~\ref{table:stateofart}.}.
Regarding the scheme, the different axes were classified
according to Tables~\ref{tab:consistency}, \ref{tab:symmetry},
\ref{tab:update}, and \ref{tab:traceability}.  Regarding the semantic
properties, for every property of Table~\ref{tab:properties}, we use
right arrows to denote that the property holds for \ensuremath{\mathsf{to}} and a left arrow
for \ensuremath{\mathsf{from}}. A normal arrow denotes that a property is satisfied by all well-behaved BXs in a given approach, while
a dashed arrow signals that well-behaved BXs only satisfy a weaker version. The absence
of an arrow means that well-behaved BXs do not satisfy a particular property, or that such property is not explicitly stated by the authors or implied by other
properties. This also means that laws implied by others are not depicted. For instance, as we have seen, correctness and hippocraticness in lenses degenerates into invertibility and stability, due to the consistency relation \ensuremath{\Varid{b}\;\mathsf{R}\;\Varid{a}\equiv \Varid{b}\mathrel{=}\mathsf{to}\;\Varid{a}}.
For totality, the absence of an arrow means that the
transformation is partial, and a dashed arrows means that the
transformation is safe.

Some particular entries in Table~\ref{table:stateofart} do not represent concrete BX tools but proposals of BX frameworks. 
Such entries are signaled with an asterisk \texttt{*}: marked properties indicate the criteria for well-behaved BXs in such frameworks. The duplicate entry of~\cite{Meertens:98} is due to the fact that two different approaches are presented in it.

Table~\ref{table:stateofart} is not intended to be complete, but rather to provide a high-level picture of existing BX tools.
It must be read with some caution, though,
since it does not capture specific intricacies of particular
frameworks that are not representable in our generic scheme, and since
some of them are omissive or ambiguous regarding particular features,
which leads to some subjectivity in the classification.  To alleviate
this we intend to publish the current survey online, and engage the
authors in the classification of their own tools and frameworks. With
completeness in mind, we also appreciate any suggestion of additional
classification axes and BX frameworks to include in the survey, in
order to reach a detailed global picture of the state of the art of
the field.

\section{Related work}
\label{section:relwork}

Acknowledging the heterogeneity of the field,~\cite{GRACE:09} surveys the related literature on BXs, grouping existing work by subcommunities, identifying some of the grand challenges of the field and providing a modest discussion on the terminology, key concepts and semantic properties used across the represented communities. 
A more focused picture on bidirectional model transformations is given in~\cite{Stevens:2008b}, with emphasis on tool support and inherent open challenges. 
Acknowledging the growing effort in the BX community towards unification, \cite{Terwilliger:2012} compares five particular BX tools in terms of their strengths and weaknesses for particular scenarios. For such comparison, it proposes a simple taxonomy 
to analyze the behavior of BX tools from a model-driven perspective, but that does not consider particular BX properties.

A detailed feature model for the classification of model transformation approaches is proposed in \cite{Czarnecki:06}, together with a survey of a vast number of existing approaches.
Nevertheless, this classification is focused on design features rather than semantic properties, and does not pay particular detail to bidirectionality.
A detailed scheme for classifying a wide spectrum of bidirectional model synchronization axiomatizations and features is provided in \cite{Antkiewicz:2008}, illustrating particular instantiations with examples of existing systems. Still, only design properties are considered, and not the semantic properties of each instance.
In contrast, \cite{Diskin:08} proposes a classifying system in which some state-based BX frameworks can be compared and analyzed in terms of their semantic laws. Despite most laws proposed there for lenses, maintainers and trigonal systems matching our own instantiations (see Section~\ref{sec:properties}), invertibility and undoability do not. In fact, the laws for maintainers assume some surjectivity constraints on the transformations, while those for trigonal systems do not. This is an indication that no consistent method for the instantiation of the laws was followed.
A general framework for building delta-based model synchronization frameworks is proposed in~\cite{DiskinGTTSE:11}, which
takes the shape \ensuremath{\overrightarrow{\mathsf{U}}\mathrel{=}\overleftarrow{\mathsf{U}}\mathrel{=}\text{D}} and \ensuremath{\overrightarrow{\mathsf{T}}\mathrel{=}\overleftarrow{\mathsf{T}}\mathrel{=}\text{D}} and considers generic laws similar to those from Table~\ref{tab:properties}, except for invertibility and least-update. However, it is not explored how existing frameworks can be instantiated under this general framework, and how the choices made on the scheme affect the properties. We follow a bottom-up characterization of the laws from state-based mappings to symmetric delta lenses, and also provide an extensive comparison of existing BX tools.

\section{Conclusions and future work}
\label{section:conclusion}

In this paper, we have presented a generic scheme and several generic
properties of BX frameworks. The generic scheme can be instantiated
along two main axes (update and traceability representation), and the
presented properties cover the most for bidirectional laws proposed in
the literature. We have also shown how this generic presentation can
be instantiated to obtain and compare most of the existing concrete BX
frameworks, such as lenses or maintainers. We have applied
this comparative study not only to such broad framework categories,
but to a large set of concrete BX tools and techniques proposed in the
literature. In the future we intend to
extend this survey effort with more entries and classification axes (covering, for
example, deployment features, such as data domain or
bidirectionalization technique) to achieve a detailed global picture
of the state of the art of the field.

One of the key roles of properties is to ensure some degree of
predictability concerning the behavior of BXs. As seen in our survey, most frameworks only guarantee
stability, correctness and invertibility, but unfortunately these
properties still leave a lot of room for unpredictable and sometimes
unreasonable behavior, defeating their goal as a means for comparing
the effectiveness of two BX frameworks. In the long term,
this problem should be solved with the study of new properties
that better characterize more refined behavior, like minimization of update translation. Meanwhile, in the
continuation of~\cite{GRACE:09}, we intend to address the problem by
developing a suite of paradigmatic examples for each application
domain, so that the user can compare the behavior of different
frameworks within that domain.
Moreover, citing~\cite{Terwilliger:2012}:
\begin{quote}
	A more ambitious goal would be a truly unified theoretical foundation to BX. [...]
	Such a unification would not be trivial to accomplish, since it will require a huge collaborative effort, involving researchers from distinct communities, countries and scientific cultures. However, such a unification would be desirable, both for better addressing existing bidirectional scenarios and for tackling largely unexplored, yet important scenarios.
\end{quote}
We believe our work presents a solid base towards such unified theoretical foundation. In its current form, we think it might help developers when designing new frameworks and end-users when comparing existing ones. Finally, we intend to improve the
accuracy and completeness of our survey by enrolling the 
BX community: the taxonomy and current results
will be published online, and authors will be invited to discuss and
extend it by classifying their own frameworks.

\section*{Acknowledgements}
This work is funded by ERDF - European Regional Development Fund through the COMPETE Programme (operational programme for competitiveness) and by national funds through the FCT - Funda{\c c}{\~ a}o para a Ci{\^ e}ncia e a Tecnologia (Portuguese Foundation for Science and Technology) within project FCOMP-01-0124-FEDER-020532. The first author of the paper is also sponsored by FCT grant SFRH/BD/69585/2010.

\bibliographystyle{splncs03}
\bibliography{survey}

\end{document}